\documentclass[twocolumn]{aastex631}

\usepackage{graphicx}	
\usepackage{amsmath}	
\usepackage{amssymb}	
\usepackage{threeparttable}
\usepackage{multirow}
\usepackage{algorithm}
\usepackage{algpseudocode}
\defcitealias{2015MNRAS.451.1728D}{DG15}

\shorttitle{A multi-wavelength SED model}
\shortauthors{Qiu \& Kang}
\graphicspath{{./}{./}}

\begin{document}

\title[A multi-wavelength SED model]{Starduster: A multi-wavelength SED model based on radiative transfer simulations and deep learning}

\correspondingauthor{Yisheng Qiu}
\email{yishengq@zju.edu.cn}
\correspondingauthor{Xi Kang}
\email{kangxi@zju.edu.cn}

\author[0000-0002-7716-1094]{Yisheng Qiu}
\affiliation{
Zhejiang University–Purple Mountain Observatory Joint Research Center for Astronomy, Zhejiang University, Hangzhou 310027, China} 
\author[0000-0002-5458-4254]{Xi Kang}
\affiliation{
Zhejiang University–Purple Mountain Observatory Joint Research Center for Astronomy, Zhejiang University, Hangzhou 310027, China} 
\affiliation{
Purple Mountain Observatory, 10 Yuan Hua Road, Nanjing 210034, China}
\affiliation{National Basic Science Data Center, Zhongguancun South 4th Street, Beijing 100190, China}



\begin{abstract}
We present \textsc{starduster}, a supervised deep learning model that predicts the multi-wavelength SED from galaxy geometry parameters and star formation history by emulating dust radiative transfer simulations. The model is comprised of three specifically designed neural networks, which take into account the features of dust attenuation and emission. We utilise the \textsc{skirt} radiative transfer simulation to produce data for the training data of neural networks. Each neural network can be trained using $\sim 4000 - 5000$ samples. Compared with the direct results of the \textsc{skirt} simulation, our deep learning model produces $\sim 0.005$ mag and $\sim 0.1 - 0.2$ mag errors for dust attenuation and emission respectively. As an application, we fit our model to the observed SEDs of IC4225 and NGC5166. Our model can reproduce the observations, and provide reasonable measurements of the inclination angle and stellar mass. However, some predicted geometry parameters are different from an image-fitting study. Our analysis implies that including a constraint at (rest-frame) $\sim 40 \, \micron$ could alleviate the degeneracy in the parameter space for both IC4225 and NGC5166, leading to broadly consistent results with the image-fitting predictions. Our SED code is publicly available and can be applied to both SED-fitting and SED-modelling of galaxies from semi-analytic models.
\end{abstract}



\section{Introduction}
Theoretical spectral energy distribution (SED) models play an important role in understanding galaxy formation. They have two main applications. The models can be fit to an observed SED to infer galaxy parameters such as dust mass, stellar mass and star formation rate \citep[e.g][]{2008MNRAS.388.1595D,2014ApJS..215....2H,2017ApJ...837..170L}. Secondly, we can also apply the models to producing large samples of mock SEDs using galaxy parameters obtained from simulations, and compare the derived statistical properties such as luminosity functions and colour-magnitude relations with observations \citep[e.g.][]{2019MNRAS.489.1357Q,2019MNRAS.483.2983Y,2020MNRAS.492.5167V}. The two applications use the theoretical SED models inversely, and can therefore be put into the same framework, leading to a better connection between observations and simulations. For example, the SED code developed by \cite{2020MNRAS.495..905R} was applied to both SED-fitting \citep{2020MNRAS.498.5581B} and SED-modelling of galaxies from a semi-analytic model \citep{2019MNRAS.489.4196L,2020MNRAS.499.1948L}.
\par
There is a gap in the comparisons between observations and simulations via theoretical SED models. Currently, most SED models use phenomenological parameters to calculate dust attenuation and emission, e.g. attenuation curve slope and dust emission spectral index, which are not direct predictions of galaxy formation simulations. This problem is typically addressed by employing empirical relations, which link dust parameters to some relevant galaxy properties \citep[e.g.][]{2017MNRAS.470.3006C,2019MNRAS.489.1357Q,2021arXiv210609741H}. However, the extra degrees of freedom introduced by the empirical relations weaken the constraining power of observations, leading to a difficult interpretation of the results. For instance, by adjusting the free parameters in an empirical dust model, \cite{2021arXiv210609741H} found that the UV and optical colour-magnitude relations can be reproduced by three hydrodynamical simulations that predict different intrinsic galaxy properties. An additional problem is that multi-wavelength SED models \citep[e.g][]{2008MNRAS.388.1595D,2017ApJ...837..170L,2019A&A...622A.103B,2020MNRAS.495..905R} assume the same dust attenuation curve to estimate both the UV to optical attenuation and the energy absorbed by dust. This is unrealistic, since the dust emission spectrum does not depend on inclination angle, while the UV to optical attenuation does. \cite{2021ApJ...923...26D} found that including inclination dependence in SED-fitting affects the estimation of stellar mass.
\par
The above issues can be overcome using dust radiative transfer simulations (e.g. \citealp{2019A&A...623A.143F,2020A&C....3100381C,2021ApJS..252...12N}, see \citealp{2013ARA&A..51...63S} for a review). These simulations estimate dust attenuation and emission by solving the dust radiative transfer equation given the star-dust geometry and dust extinction curves. The dependence of the dust attenuation on inclination angle is explicitly taken into account by these simulations. However, they are computationally expensive, and therefore are not widely used in SED-fitting or producing large mock galaxy catalogues \citep[however see][]{2018ApJS..234...20C}. 
\par
The main purpose of this work is to develop a supervised deep learning model that emulates radiative simulations. Our model is trained using data generated by radiative simulations, and predicts the FUV to FIR SED from galaxy geometry parameters and star formation history. As complementary, some studies build the model from the opposite direction, i.e. deriving the galaxy and dust properties from a SED. Such models can be found in \cite{2019MNRAS.490.5503L,2020A&A...634A..57D,2021ApJ...916...43G}. This work also gives an example of using the deep learning model in SED-fitting. An application of our SED model to a semi-analytic model will be presented in future work.
\par
This paper is organised as follows. Section \ref{sec:method} describes our deep learning model. Section \ref{sec:app} demonstrates an application of the model to SED-fitting. Finally, this work is summarised in Section \ref{sec:summary}. This paper uses concepts from deep learning directly, e.g. the fully connected layer and the multi-layer perceptron. The reader is referred to \cite{Goodfellow-et-al-2016} for a nice introductory book on deep learning. 
\par
Our code, named as \textsc{starduster}, is available on GitHub\footnote{\url{https://github.com/yqiuu/starduster}} under a GPLv3 License and version 0.1.0-beta is archived in Zenodo \citep{starduster}. The implementation of the deep learning model includes many technical details and hyperparameters. It is tedious to present all of them. The reader is referred to our public code repository for the details.

\section{Methodology} \label{sec:method}

\begin{table*}
    \movetableright -15mm
	\centering
	\caption{Summary of the input parameters to run the \textsc{skirt} dust radiative transfer simulation.}
	\label{tab:params}
	\begin{tabular}{cccc} 
		\hline
		Symbol & Description & Value & Unit \\
		\hline
		$\theta$ & Inclination angle & Beta distribution with $\alpha = \beta = 0.5$ $^\text{a}$ & - \\
		$r_\text{disk}$ & Stellar disk radius & Log-uniform distribution from -2.0 to 2.5 & kpc \\
		$h_\text{disk}/r_\text{disk}$ & Stellar disk height scaled by stellar disk radius & Fixed at 0.1 & - \\
		$r_\text{bulge}$ & Stellar bulge radius & Log-uniform distribution from -0.5 to 1.5 & kpc \\
		$r_\text{dust}/r_\text{disk}$ & Dust disk radius scaled by stellar disk radius & Log-uniform distribution from -0.7 to 0.7 & - \\
		$h_\text{dust}/r_\text{disk}$ & Dust disk height scaled by stellar disk radius &  Fixed at 0.1 & - \\
		$\Sigma_\text{dust}$ & Dust column density$^\text{b}$ & Log-uniform distribution from 3.5 to 7.5 & $\text{M}_\odot/\text{kpc}^2$ \\
		\hline
		$\eta^\text{tot}$ & Intrinsic bolometric luminosity & Log-uniform distribution from 6 to 14 & $\text{L}_\odot$ \\
		$\alpha_\text{B/T}$ & Intrinsic Bulge to total luminosity ratio & Beta distribution with $\alpha = \beta = 0.5$ & - \\
		\hline
	\end{tabular}
	\begin{tablenotes}
        \item $^\text{a}$ Inclination angle is scaled to [0, $\frac{\pi}{2}$] using linear transformation.
        \item $^\text{b}$ Dust column density is defined by $\Sigma_\text{dust} = m_\text{dust} / (2\pi r^2_\text{dust})$.
    \end{tablenotes}
\end{table*}

\subsection{Geometry model} \label{sec:geometry}
We start by describing the geometry model of a galaxy, which serves as the input of radiative transfer simulations. We assume that a galaxy is comprised of a stellar disk, a stellar bulge and a dust disk. Their density profiles are given by the following parametric forms:
\begin{align}
    &\rho^\text{disk}(r, z) \propto \exp \left(- \frac{r}{r^\text{disk}} - \frac{|z|}{h^\text{disk}}\right), \label{eqn:rho_disk} \\ 
    &\rho^\text{bulge}(r, z) \propto \mathcal{S}_n \left(\frac{\sqrt{r^2 + z^2}}{r^\text{bulge}} \right), \label{eqn:rho_bulge} \\
    &\rho^\text{dust}(r, z) \propto \exp \left(- \frac{r}{r^\text{dust}} - \frac{|z|}{h^\text{dust}}\right), \label{eqn:rho_dust}
\end{align}
where $\mathcal{S}_n$ is the S\'ersic function:
\begin{align}
    & \mathcal{S}_n(s) = -\frac{1}{\pi} \int^\infty_s \frac{dI}{dt} \frac{dt}{\sqrt{t^2 - s^2}}, \\
    & I(t) = \frac{b^{2n}}{\pi \Gamma(2n + 1)} \exp(-bt^{1/n}).
\end{align}
This geometry model mainly follows \cite{2013A&A...550A..74D,2019A&A...623A.143F}, which are frequently used to study observed spiral galaxies \citep[e.g.][]{2014A&A...571A..69D,2014MNRAS.441..869D,2015MNRAS.451.1728D,2017A&A...599A..64V}.

\subsection{The dust attenuation model}
\subsubsection{Neural network design} \label{sec:nn_da}

\begin{figure*}
	\includegraphics[width=\textwidth,trim=0 60 0 10,clip]{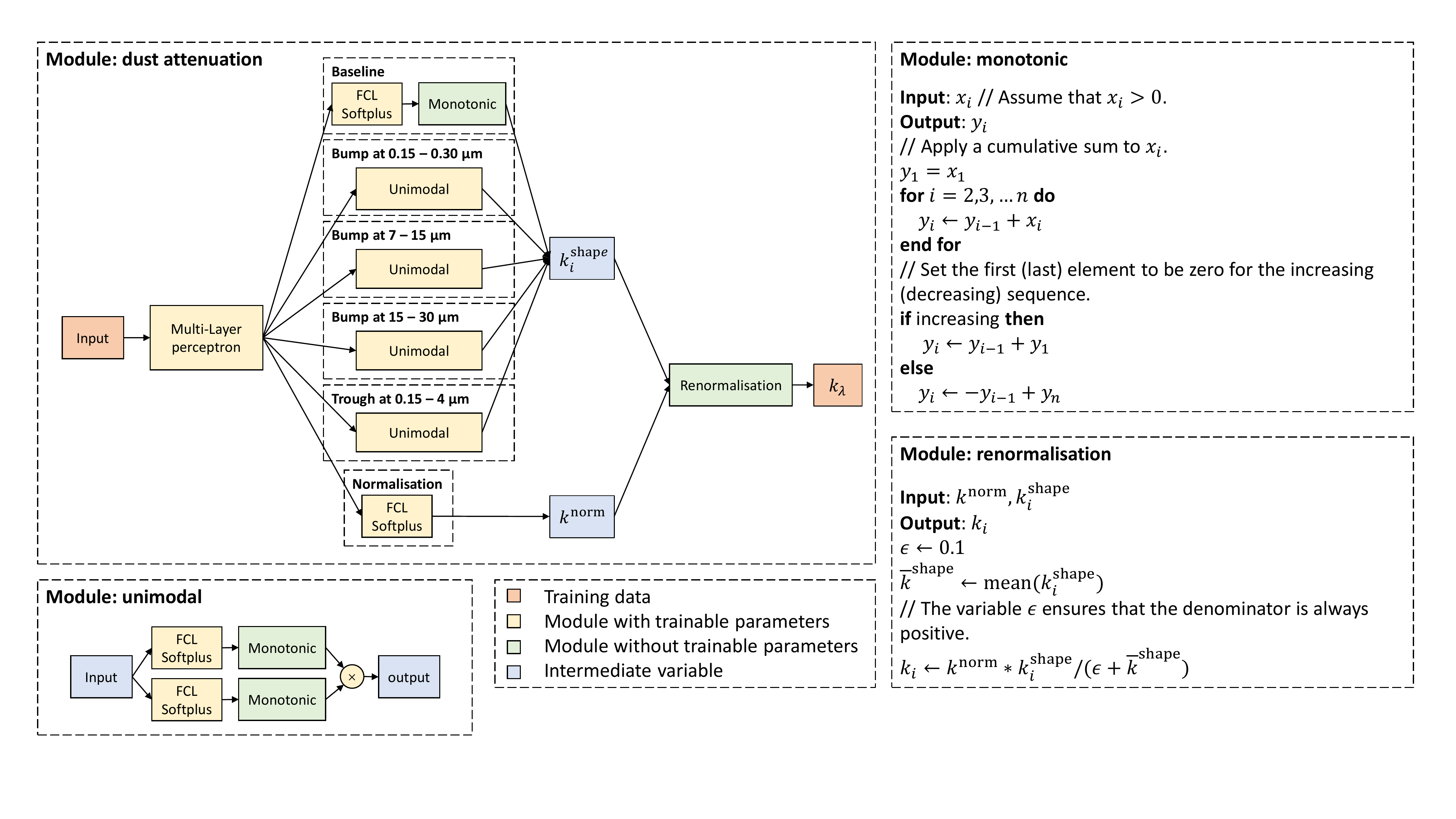}
    \caption{Schematic diagram of the dust attenuation neural network described in Section \ref{sec:nn_da}. The neural network is designed to explicitly take into account the features of the attenuation curves generated by the \textsc{skirt} radiative simulation. We suggest that these curves can be decomposed into five components, i.e. a monotonically decreasing baseline, a trough, and three bumps. These features can be seen in Figure \ref{fig:curves}.}
    \label{fig:nn_da}
\end{figure*}

\begin{figure}
    \centering
	\includegraphics[width=\columnwidth]{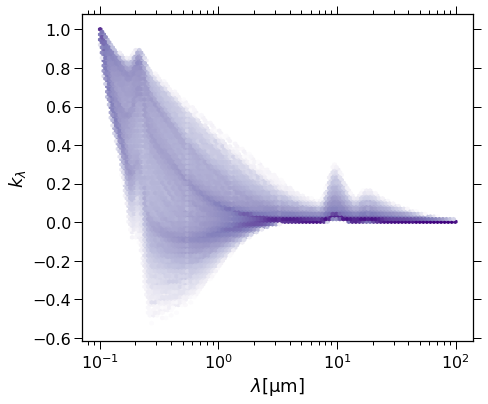}
    \caption{Heat map of dust attenuation curves produced by the \textsc{skirt} radiative simulation. These curves are generated in Section \ref{sec:data_da}, which are used to train the dust attenuation neural work of the stellar disk. The curves are such normalised that the value at the shortest wavelength is equal to one.}
    \label{fig:curves}
\end{figure}

Our dust attenuation model is designed to compute the dust obscured stellar luminosity. With the geometry model described in the previous section and the linearity of the radiative transfer equation, we can express the dust attenuated luminosity as
\begin{equation}
    L^\text{star}_\lambda = 10^{-0.4 k^\text{disk}_\lambda} L^\text{disk}_\lambda +  10^{-0.4 k^\text{bulge}_\lambda} L^\text{bulge}_\lambda, \label{eqn:k_lam}
\end{equation}
where $L^\text{disk}_\lambda,L^\text{bulge}_\lambda$ are the intrinsic luminosities, and $k^\text{disk}_\lambda,k^\text{bulge}_\lambda$ are the corresponding dust attenuation curves. We note that the dust attenuation curves can be obtained by simulations that contain only one stellar component, which can be run separately and are independent of the intrinsic luminosity. Accordingly, our approach first carries out two sets of simulations that predict $k^\text{disk}_\lambda$ and $k^\text{bulge}_\lambda$, and then uses the resulting data to train two neural networks.
\par
The design of the proposed neural network is illustrated in Figure \ref{fig:nn_da}. To understand the motivation of this design, we show some attenuation curves of the disk component as a heat map in Figure \ref{fig:curves}. These data are produced in the next section. In order to highlight the shape, these curves are normalised so that the value at the shortest wavelength is equal to one. We suggest that these curves can be decomposed into five components, i.e. a monotonically decreasing baseline, a trough, and three bumps. Our neural network takes into account these features explicitly. We can construct a sequence of data that have one of these features in the following way. Firstly, a monotonically increasing sequence can be obtained by applying a cumulative sum to an array of positive data. Then, by changing the sign, we can create a monotonically decreasing sequence. Finally, we can generate a sequence that has only one peak or trough by multiplying a monotonically increasing sequence and a monotonically decreasing sequence together. We introduce a renormalisation module as a final step in the neural network. This step ensures those modules which represent the features of the curves have similar scales.

\subsubsection{Data preparation} \label{sec:data_da}
We utilise the \textsc{skirt} 3D Monte Carlo dust radiative transfer simulation \citep{2011ApJS..196...22B,2015A&C.....9...20C,2020A&C....3100381C} to generate training and validation data. Specifically, we use version 9 of \textsc{skirt} introduced in \cite{2020A&C....3100381C}.
\par
We run the \textsc{skirt} radiative transfer simulation using the geometry model described in Section \ref{sec:geometry}. In addition to those defined in equation (\ref{eqn:rho_disk}) - (\ref{eqn:rho_dust}), the input parameters also include dust mass and inclination angle. We sample these parameters according to Table \ref{tab:params}.
\par
In addition, we adopt the \cite{2007ApJ...657..810D} dust grain mixture, and launch $10^7$ photon packages for each run. The dust emission luminosity is not calculated for these simulations, since we are only interested in the dust attenuation curves. The output wavelength grid ranges from $0.09 \, \mu \text{m}$ to $100 \, \mu \text{m}$, with 300 bins in logarithmic scale.
\par
Data preprocessing is an essential step in the machine learning context. We scale all input parameters into $[-1, 1]$. Specifically, we apply $f(\theta) = 2\cos(\theta) - 1$ for the inclination angle, and $f(x) = 2\frac{\log x - \log x_\text{min}}{\log x_\text{max} - \log x_\text{min}} - 1$ for other parameters. The output curves are noisy due to the Monte Carlo nature of the simulations. To reduce the noise, we smooth the attenuation curves using a Savitzky-Golay filter. Furthermore, we exclude the data for $k(\lambda = 100 \, \micron) < 0.01$, since they are extremely noisy and badly behaved. Table \ref{tab:number} shows the number of samples in each data set.

\subsubsection{Training} \label{sec:training_da}
We adopt the following loss function for the dust attenuation neural networks
\begin{equation}
    \Psi = \sum_\text{batch} \max_\lambda \left| {k^\text{pred}_\lambda} - {k^\text{true}_\lambda} \right|.
\end{equation}
To minimise the loss function, we employ an Adam optimiser with a batch size of 256. We run the optimiser for 3000 epochs and change the learning rate dynamically according to \citet{2019SPIE11006E..12S}. We train the neural network four times with different maximum learning rates at $2.5 \times 10^{-4}$, $5 \times 10^{-4}$, $1 \times 10^{-3}$ and $2 \times 10^{-3}$, and choose the best result from these runs.
\par
\begin{figure*}
	\includegraphics[width=\textwidth]{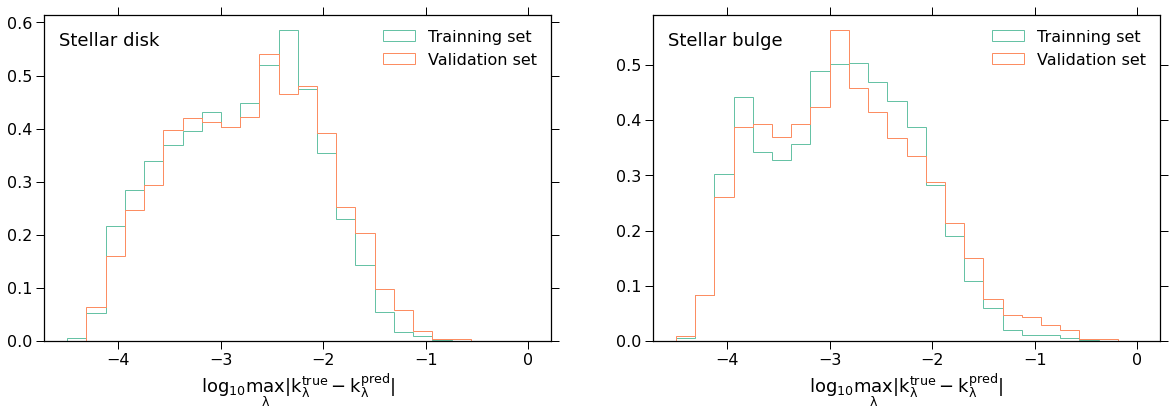}
    \caption{Error distributions of the trained dust attenuation models. Left and right panels correspond to the models of stellar disk and bulge respectively. The training approach is described in Section \ref{sec:training_da}.}
    \label{fig:loss_da}
\end{figure*}

Figure \ref{fig:loss_da} illustrates the training results. The median error on the dust attenuation curves is roughly $3 \times 10^{-3}$ mag for both the stellar disk and bulge models. We will discuss the uncertainties in more detail in Section \ref{sec:valid}, where we combine the dust attenuation and emission neural networks.

\subsubsection{Classifiers} \label{sec:exclude}
As mentioned in Section \ref{sec:data_da}, we exclude some badly behaved samples from the training data, and consequently the obtained neural networks should not be applied to the excluded parameter space. For this reason, we train two classifiers for the disk and bulge models respectively to identify whether the input parameters are in the effective region. Each classifier is a multi-layer perceptron with one hidden layer. Both classifiers are trained using the same method as in Section \ref{sec:training_da}.

\begin{table}
    \movetableright -20mm
	\centering
	\caption{Number of samples in the data set generated by this work.}
	\label{tab:number}
	\resizebox{1.05\columnwidth}{!}{
	\begin{tabular}{ccccc} 
		\hline
		Data set & Total & Included & Training & Validation \\
		\hline
        Dust attenuation (disk) & 7020 & 5850 & 4094 & 1756 \\
        Dust attenuation (bulge) & 7020 & 6007 & 4204 & 1803 \\
        Dust emission & 8100 & 6772 & 4740 & 2032 \\
        Integrated validation & 2048 & - & - & 2048 \\
        \hline
	\end{tabular}}
\end{table}

\subsection{The dust emission model}
\subsubsection{Neural network} \label{sec:nn_de}

\begin{figure*}
	\includegraphics[width=\textwidth,trim=0 60 0 10,clip]{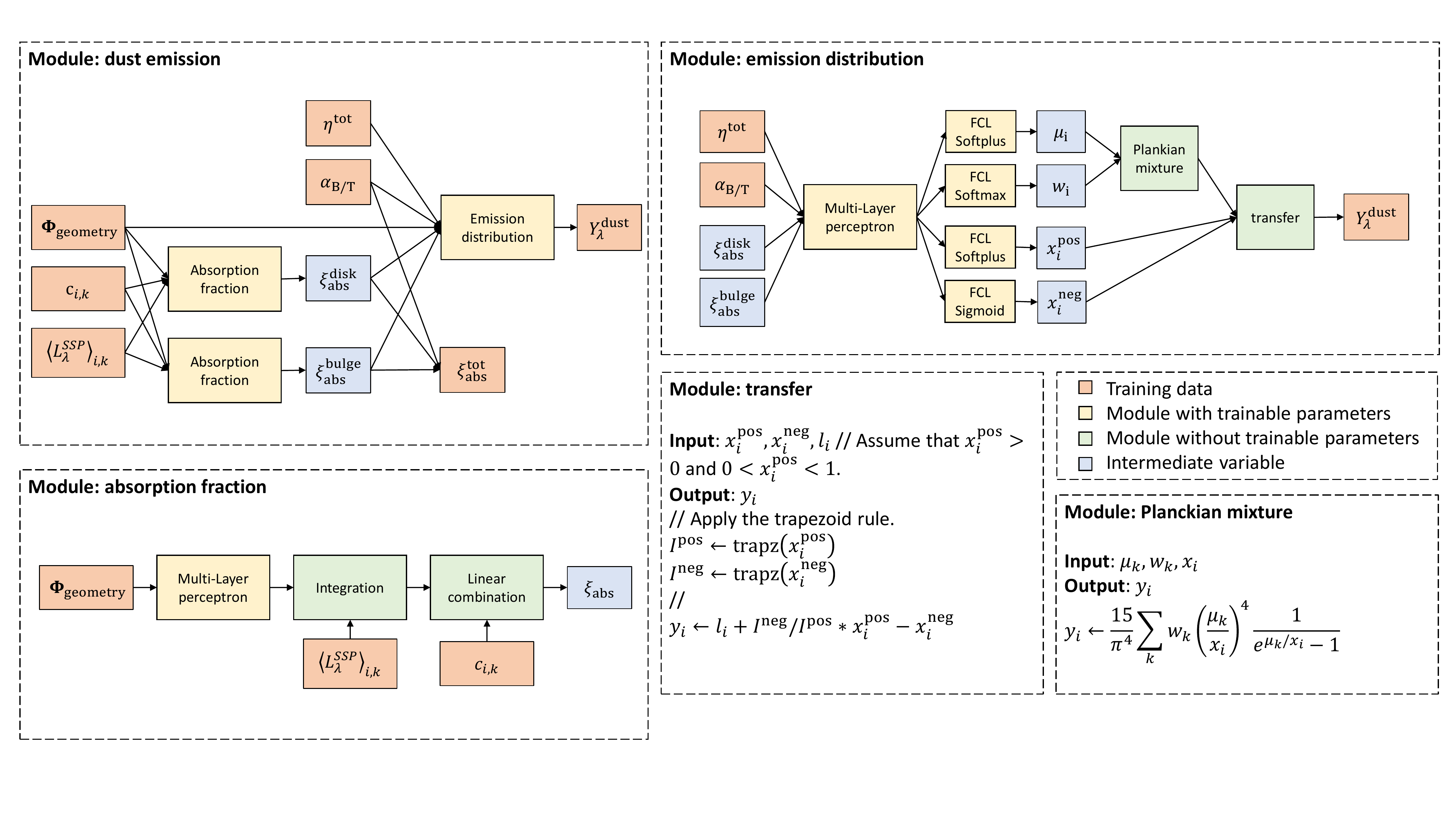}
    \caption{Schematic diagram of the proposed dust emission neural network described in Section \ref{sec:nn_de}. The model assumes a discrete star formation history parameterised by $c_{i,k}$. While the neural network works for any $c_{i,k}$, its training by design can be done without sampling $c_{i,k}$. In addition, the model uses a Planckian mixture to model the FIR dust emission and a transfer module to describe the MIR dust emission.}
    \label{fig:nn_de}
\end{figure*}

Our dust emission neural network aims to predict the dust emission luminosity. A major difficulty is that the dependence of the dust emission luminosity on the intrinsic SED is non-linear, and therefore the training data must include the intrinsic SED. While the intrinsic SED can be derived from star formation history and a simple stellar population library, it is almost impossible to fully sample star formation history. Fortunately, we have recognised that the dust emission luminosity depends on the intrinsic SED only via the dust absorption fraction, which is a linear function with respect to the intrinsic SED. This is the key point behind the design of our dust emission neural network. In general, the dust emission luminosity also relies on dust compositions, and in this work, we assume a universal dust composition.
\par
The proposed dust emission neural network is illustrated in Figure \ref{fig:nn_de}. To understand our approach, we write the dust emission luminosity as
\begin{equation}
    \mathcal{L}^\text{dust}_\lambda = \eta^\text{tot} \xi^\text{tot}_\text{abs} Y^\text{dust}_\lambda,
\end{equation}
with
\begin{equation}
    \eta^\text{tot} = \int L^\text{disk}_\lambda + L^\text{bulge}_\lambda \, d\lambda,
\end{equation}
where $\eta^\text{tot}$ is the intrinsic bolometric luminosity, $\xi^\text{tot}_\text{abs}$ is the dust absorption fraction, and $Y^\text{dust}_\lambda$ is the normalised luminosity. We first focus on the modelling of the absorption fraction. The linearity of the energy balance equation suggests that the absorption fraction can be decomposed into two parts
\begin{equation}
    \xi^\text{tot}_\text{abs} = (1 - \alpha_\text{B/T}) \xi^\text{disk}_\text{abs} + \alpha_\text{B/T} \xi^\text{bulge}_\text{abs},
\end{equation}
with
\begin{equation}
    \alpha_\text{B/T} = \frac{1}{\eta^\text{tot}} \int L^\text{bulge}_\lambda d \lambda,
\end{equation}
where $\xi^\text{disk}_\text{abs}$ and $\xi^\text{bulge}_\text{abs}$ are the fractions of energy contributed by the stellar disk and bulge but absorbed by dust respectively, and $\alpha_\text{B/T}$ is the intrinsic bulge to total luminosity ratio. We suggest that $\xi^\text{disk}_\text{abs}$ and $\xi^\text{bulge}_\text{abs}$ can have the following functional forms
\begin{align}
    \xi^\text{disk}_\text{abs} &= \int \mathcal{A}^\text{disk}_\lambda  Y^\text{disk}_\lambda \, d\lambda, \label{eqn:xi_disk} \\
    \xi^\text{bulge}_\text{abs} &= \int \mathcal{A}^\text{bulge}_\lambda Y^\text{bulge}_\lambda \, d\lambda, \label{eqn:xi_bulge}
\end{align}
where $\mathcal{A}^\text{disk}_\text{abs}$ and $\mathcal{A}^\text{bulge}_\text{abs}$ are the mean dust attenuation functions. They can be modelled using neural networks. Different from the dust attenuation curves defined in Equation (\ref{eqn:k_lam}), these two functions are independent of inclination angle. In principle, they can be derived from $k^\text{disk}_\lambda$ and $k^\text{bulge}_\lambda$. However, this approach is not adopted in this work, since it introduces an intermediate training step, resulting in extra errors. Instead, we make the dust absorption fractions linearly depend on the star formation history parameters\footnote{The star formation history parameters defined here are different from those used in SED-fitting. They are two-dimensional and depend on both stellar age and metallicity.} by using discrete star formation history, in which case the normalised intrinsic SED can be written as 
\begin{equation} \label{eqn:ssp_1}
    Y_\lambda = \sum_{i}^{n_Z} \sum_{k}^{n_\tau} c_{i,k}
    {\langle Y^\text{SSP}_\lambda \rangle}_{i,k},
\end{equation}
with
\begin{equation} \label{eqn:ssp_2}
 {\langle Y^\text{SSP}_\lambda \rangle}_{i,k} = \int^{t_{k+1}}_{t_{k}} Y^\text{SSP}_\lambda(\tau, Z_i) \, d\tau,
\end{equation}
where $c_{i,k}$ is the fractional luminosity in $i$-th metallicity and $k$-th stellar age bins, and $Y^\text{SSP}_\lambda(\tau, Z)$ is the normalised spectrum of simple stellar population (SSP) with stellar age $\tau$ and metallicity $Z$. Substituting equations (\ref{eqn:ssp_1}) and (\ref{eqn:ssp_2}) into equations (\ref{eqn:xi_disk}) and (\ref{eqn:xi_bulge}) yields
\begin{align}
    & \xi^\text{disk}_\text{abs} = \sum_{i}^{n_Z} \sum_{k}^{n_\tau} c^\text{disk}_{i,k} \zeta^\text{disk}_\text{abs}, \\
    & \xi^\text{bulge}_\text{abs} = \sum_{i}^{n_Z} \sum_{k}^{n_\tau} c^\text{bulge}_{i,k} \zeta^\text{bulge}_\text{abs}, \label{eqn:c_bulge}
\end{align}
with
\begin{align}
    & \zeta^\text{disk}_\text{abs} = \int \mathcal{A}^\text{disk}_\lambda {\langle Y^\text{SSP}_\lambda \rangle}_{i,k} \, d\lambda, \\
    & \zeta^\text{bulge}_\text{abs} = \int \mathcal{A}^\text{bulge}_\lambda {\langle Y^\text{SSP}_\lambda \rangle}_{i,k} \, d\lambda,
\end{align}
The above equations suggest that once the neural network can predict $\zeta^\text{disk}_\text{abs}$ and $\zeta^\text{bulge}_\text{abs}$, the total absorption fraction can be obtained using the linear combination. Hence, when preparing the training data, there is no need to sample the star formation history of the stellar disk and bulge, i.e. $c^\text{disk}_{i,k}$ and $c^\text{bulge}_{i,k}$. Instead, these parameters can be provided by the one hot encoding. This significantly reduces the dimensions of the problem.
\par
In terms of the neural network to predict $Y^\text{dust}_\lambda$, we point out that the dependence of the dust luminosity on the intrinsic SED is only via the absorption coefficient in accordance with the energy balance equation. In other words, we can write
\begin{equation}
    Y^\text{dust}_\lambda = Y^\text{dust}_\lambda(\xi^\text{disk}_\text{abs}, \xi^\text{bulge}_\text{abs}, \mathbf{\Phi}_\text{geometry}),
\end{equation}
where $\mathbf{\Phi}_\text{geometry}$ is the parameter set describing the geometry model exclusive of the inclination angle. The functional form of $Y^\text{dust}_\lambda$ is modelled using a neural network.
\par
\begin{figure*}
	\includegraphics[width=\textwidth]{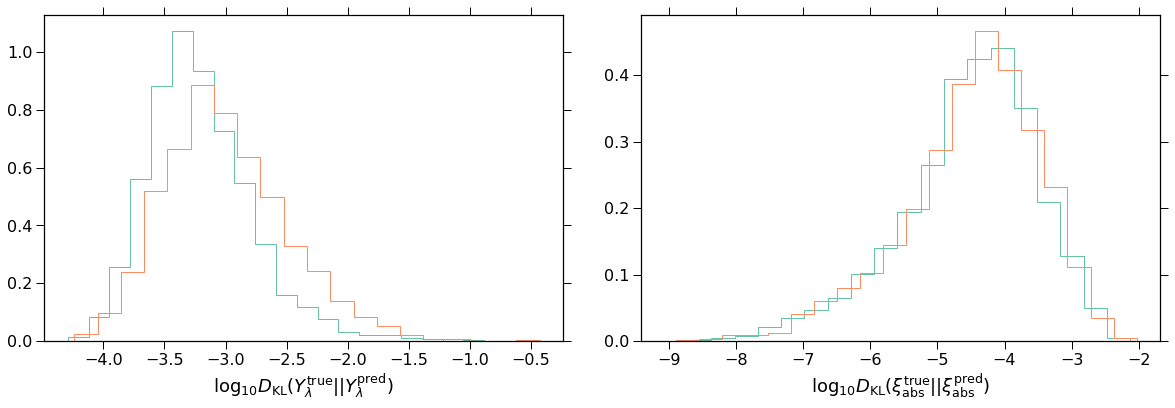}
    \caption{Error distributions of the trained dust emission model. The left and right panels show the errors of the normalised dust emission spectrum and absorption fraction respectively. The errors are defined by the KL divergences in Equations (\ref{eqn:loss_de_kl1}) and (\ref{eqn:loss_de_kl2}). The training method is described in Section \ref{sec:training_de}.}.
    \label{fig:loss_de}
\end{figure*}

As illustrated in Figure \ref{fig:loss_de}, we introduce a specific module to reproduce the shape of the dust emission luminosity. The FIR emission is modelled by a Planckian mixture, which is a weighted sum of some Planck distributions with different temperatures. The resulting spectrum is then modified to take into account the other emission features at MIR wavelengths using a transfer module. The design of the transfer module is demonstrated in Figure \ref{fig:nn_de}. The key idea behind the module is probability conservation. 

\subsubsection{Data preparation} \label{sec:data_de}
To prepare the data, we use \textsc{skirt} to perform a set of simulations with both stellar disk and bulge components. We apply the same method as in Section \ref{sec:data_da} to sample the required parameters for the simulations. Moreover, each run requires two SSP spectra as input for the disk and bulge components respectively. They are randomly chosen from a library. We construct the library using the Flexible Stellar Population Synthesis model \citep{2009ApJ...699..486C,2010ApJ...712..833C} with a \cite{2003PASP..115..763C} initial mass function. Nebular line and continuum emissions are added following \cite{2017ApJ...840...44B}. The raw SSP spectra are gridded using Equation (\ref{eqn:ssp_2}) with six stellar age bins and six metallicity bins. The simulations require two additional parameters, i.e. the intrinsic bolometric luminosity $\eta^\text{tot}$ and bulge to total luminosity ratio $\alpha_\text{B/T}$. We sample them according to Table \ref{tab:params}.
\par
For other settings of the simulations, we assume that dust grains are not in local thermal equilibrium, and use the stochastic heating algorithm to calculate the dust emission luminosity. To be consistent with the dust attenuation model, we adopt the \cite{2007ApJ...657..810D} dust grain mixture. The output wavelength grid uses 500 logarithmic bins and ranges from 0.09 $\micron$ and 10,000 $\micron$. Finally, we launch $10^7$ photon packages for each run.
\par
We adopt the same method as in Section \ref{sec:data_da} to preprocess the input parameters. In addition, we use the classifiers obtained in Section \ref{sec:exclude} to exclude badly behaved data. We find that the data exclusion can lead to better training results\footnote{In fact, this step can be done by pre-selecting the input parameters before running the simulations.}. The number of samples in the data set is given by Table \ref{tab:number}. We have tested that increasing the sample size only gives limited improvement in the training results.

\subsubsection{training} \label{sec:training_de}
To train the dust emission neural work, we define the following loss function
\begin{equation}
    \Psi = \sum_\text{batch} D_\text{KL}(Y^\text{true}_\lambda||Y^\text{pred}_\lambda) + \sum_\text{batch} D_\text{KL}(\xi^\text{true}_\text{abs}||\xi^\text{pred}_\text{abs}),
\end{equation}
with 
\begin{align}
    & D_\text{KL}(Y^\text{true}_\lambda||Y^\text{true}_\lambda) = \int Y^\text{true}_\lambda \ln{\frac{Y^\text{true}_\lambda}{Y^\text{pred}_\lambda}} \, d\lambda, \label{eqn:loss_de_kl1} \\
    & D_\text{KL}(\xi^\text{true}_\text{abs}||\xi^\text{pred}_\text{abs}) = \xi^\text{true}_\text{abs} \ln{\frac{\xi^\text{true}_\text{abs}}{\xi^\text{pred}_\text{abs}}} + (1 - \xi^\text{true}_\text{abs}) \ln{\frac{1- \xi^\text{true}_\text{abs}}{1 - \xi^\text{pred}_\text{abs}}}. \label{eqn:loss_de_kl2}
\end{align}
The normalised dust emission spectrum can be treated as a probability distribution, and it is common to quantify the distance between two probability distributions using the Kullback-Leibler (KL) divergence. The dust emission luminosity is represented as a one dimensional array in the code, and we compute the KL divergence using the trapezoid rule. Moreover, the total absorption fraction by definition ranges from 0 to 1. Accordingly, we can treat it as a probability value and quantify the difference between $\xi^\text{true}_\text{abs}$ and $\xi^\text{pred}_\text{abs}$ using the discrete KL divergence. In terms of the optimisation strategy, we adopt the same method as in Section \ref{sec:training_da}.
\par
For the best result, we calculate the KL divergences for each sample, and demonstrate the corresponding distributions in Figure \ref{fig:loss_de}. The median KL divergence is $\sim 7 \times 10^{-4}$ for the normalised emission spectra and $\sim 4 \times 10^{-5}$ for the dust absorption fraction. A more intuitive analysis of the uncertainties will be carried out in the next section, where we combine the dust attenuation and emission models. 

\begin{figure*}
	\includegraphics[width=\textwidth]{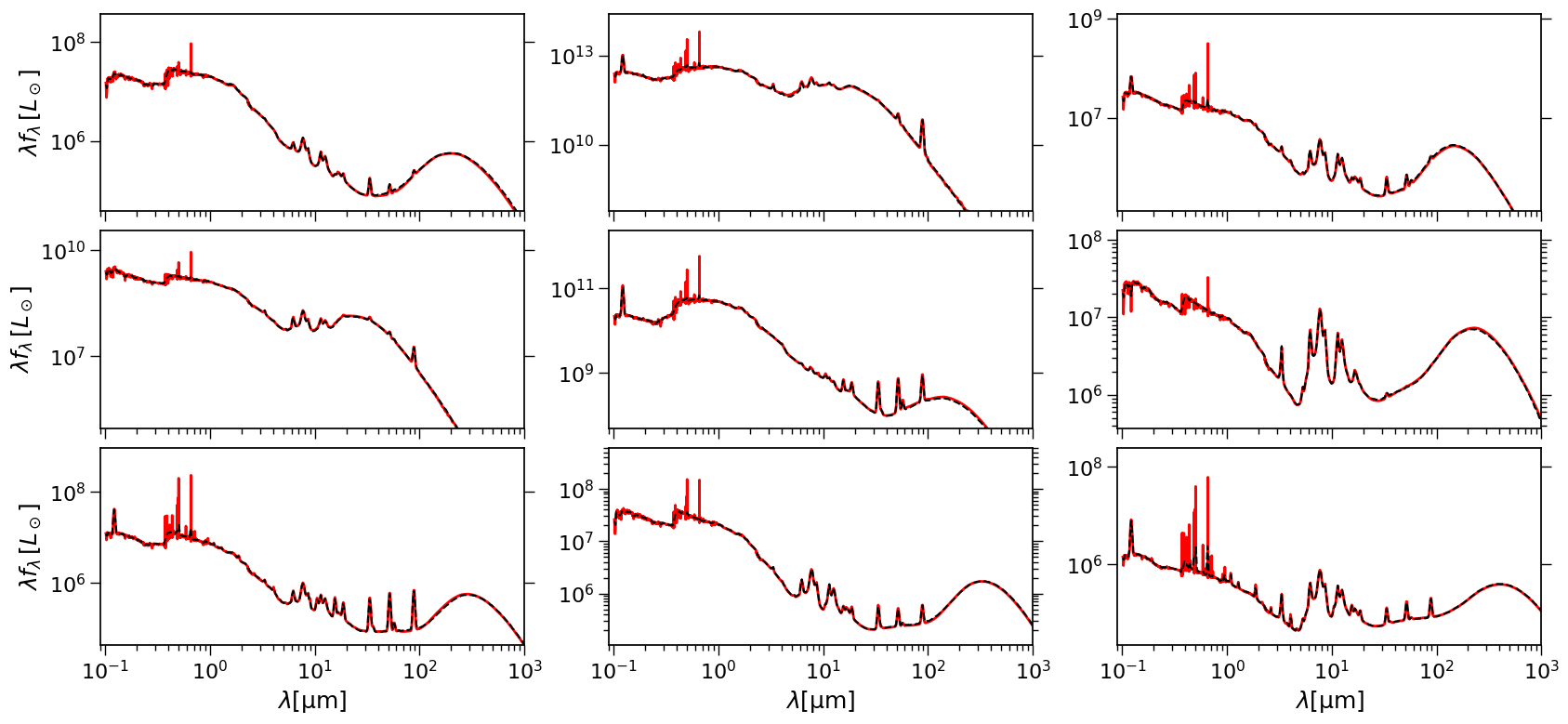}
    \caption{Example SEDs from the validation data described in Section \ref{sec:valid}. The black and red lines show the SEDs obtained by \textsc{skirt} and our deep learning model respectively.}
    \label{fig:valid_examples}
\end{figure*}

\begin{figure*}
	\includegraphics[width=\textwidth]{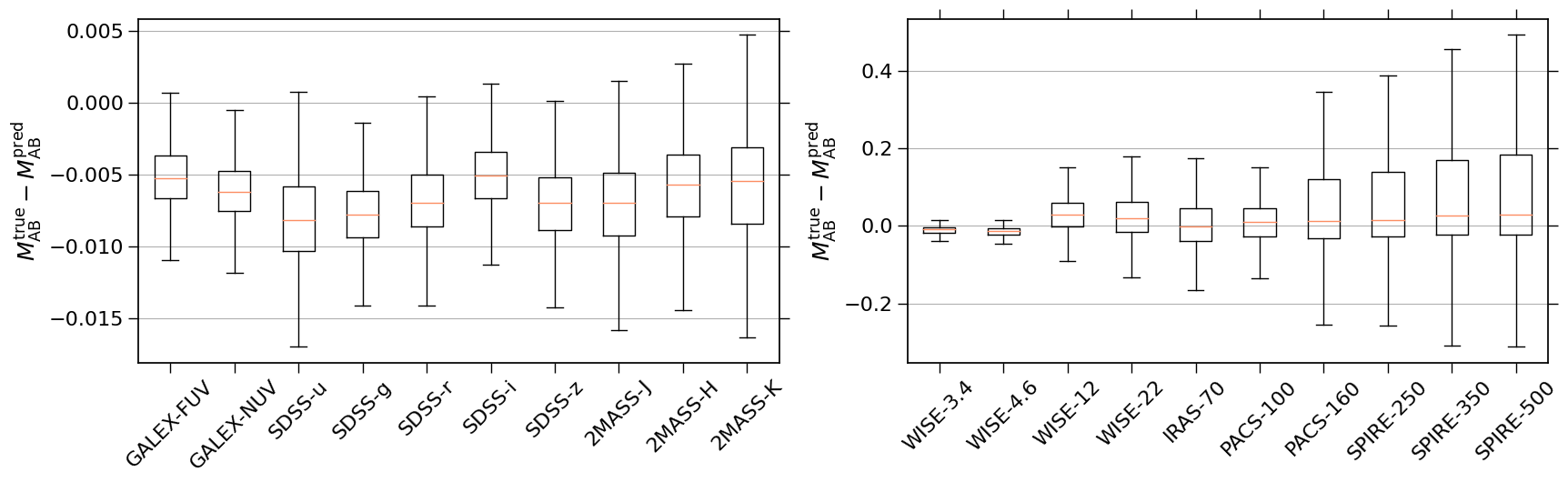}
    \caption{Uncertainties in our deep learning model in 20 filters. The model combines the dust attenuation and emission networks. The errors are given by the difference in AB magnitude and estimated using the validation data described in Section \ref{sec:valid}.}
    \label{fig:valid_errors}
\end{figure*}

\subsection{Combining dust attenuation and emission} \label{sec:valid}
In this section, we integrate the dust attenuation and emission models, and estimate the overall errors. Our multi-wavelength SED model takes the parameters listed in Table \ref{tab:params} and the star formation history parameters defined in Equations (\ref{eqn:ssp_1}) - (\ref{eqn:c_bulge}) as input. To obtain the stellar luminosity, for both the disk and bulge components, we first integrate the SSP library with the star formation history parameters and then apply the dust attenuation curves obtained by the deep learning model. Since attenuation curves are smooth, we can interpolate between them, allowing for a finer wavelength grid for the SSP library instead of the one adopted in the model training. This approach includes more emission line features from the SSP library. Dust emission luminosity is then added to the stellar luminosity to obtain a multi-wavelength SED. Linear interpolation is adopted in the calculations above.
\par
To prepare the validation data, we sample the fractional luminosity  $c^\text{disk}_{i,k}$ and $c^\text{bulge}_{i,k}$ using a Dirichlet distribution with a constant concentration parameter of 0.5. The sampling of other parameters follows Table \ref{tab:params}. To be consistent, we run the \textsc{skirt} simulation using the same configuration as in Section \ref{sec:data_da} and Section \ref{sec:data_de}. Our validation set contains 2048 samples.
\par
Figure \ref{fig:valid_examples} illustrates nine example SEDs predicted by \textsc{skirt} and our deep learning model. They are given as black and red lines respectively. We find good consistency between them. In Figure \ref{fig:valid_errors}, we demonstrate the errors in AB magnitude in twenty filters from FUV to FIR. We only compare the results for $M^\text{true}_\text{AB} < -10$. In the left panel, the errors in the bands are dominated by the dust attenuation model, which are at $\sim 0.005$ mag with a $\sim -0.006$ systematic offset. In terms of the dust emission model, as shown in the right panel, the predicted magnitudes in the MIR to FIR bands deviate from the fiducial values by $\sim 0.1$ mag to $0.2$ mag. The uncertainties in \textit{WISE}-3.4 and \textit{WISE}-4.6 are smaller at $\sim 0.03$ mag.

\begin{table*}
    \movetableright -20mm
	\centering
	\caption{Best-fitting parameters of IC4225 and NGC5166.}
	\label{tab:best-fit}
	\resizebox{\textwidth}{!}{
	\begin{tabular}{ccccccc}
		\hline
	    \multirow{2}{*}{Symbol} & \multirow{2}{*}{Description} & \multirow{2}{*}{Unit} & IC4225 & IC4225 & IC4225 & IC4225 \\
	    &&&& (fixed $r_\text{bulge}$) & (fixed $r_\text{bulge}$) & (with mock data) \\
	    \hline
		 $\theta$ & Inclination angle & - & 88.7 & 86.8 & 86.6 & 86.5 \\
		 $\Sigma_\text{dust}$ & Dust column density & $\text{M}_\odot/\text{kpc}^2$ & $1.03 \times 10^{5}$ & $6.73 \times 10^{4}$ & $6.95 \times 10^{4}$ & $7.17 \times 10^{4}$ \\
		 $r_\text{dust}/r_\text{disk}$ & Dust disk radius scaled by stellar disk radius & - & 0.83 & 4.28 & 4.24 & 4.31 \\
		 $r_\text{disk}$ & Stellar disk radius & kpc & 8.28 & 2.06 & 1.96 & 1.89 \\
		 $r_\text{bulge}$ & Stellar bulge radius & kpc & 0.32 & 0.64 & 1.25 & 1.41 \\
		 $\eta^\text{tot}$ & Intrinsic bolometric luminosity & $\text{L}_\odot$ & $4.59 \times 10^{10}$ & $4.85 \times 10^{10}$ & $4.81 \times 10^{10}$ & $4.81 \times 10^{10}$ \\
		 $\alpha_\text{B/T}$ & Intrinsic bulge to total luminosity ratio & - & 0.92 & 0.09 & 0.08 & 0.08 \\
		 $c^\text{disk}_1$ & Fractional luminosity for 1 Myr - 24 Myr & - & 1.00 & 0.00 & 0.00 & 0.00 \\
		 $c^\text{disk}_2$ & Fractional luminosity for 24 Myr - 581 Myr & - & 0.00 & 0.00 & 0.00 & 0.00 \\
		 $c^\text{disk}_3$ & Fractional luminosity for 581 Myr - 14 Gyr & - & 0.00 & 1.00 & 1.00 & 1.00 \\
		 $Z_\text{disk}$ & Disk metallicity & $\text{Z}_\odot$ & 1.33 & 0.42 & 0.42 & 0.42 \\
		 $c^\text{bulge}_1$ & Fractional luminosity for 1 Myr - 24 Myr & - & 0.00 & 0.18 & 1.00 & 1.00 \\
		 $c^\text{bulge}_2$ & Fractional luminosity for 24 Myr - 581 Myr & - & 0.00 & 0.82 & 0.00 & 0.00 \\
		 $c^\text{bulge}_3$ & Fractional luminosity for 581 Myr - 14 Gyr & - & 1.00 & 0.00 & 0.00 & 0.00 \\
		 $Z_\text{bulge}$ & Bulge metallicity & $\text{Z}_\odot$ & 0.46 & 0.002 & 0.01 & 0.009 \\
	     \hline
	    \multirow{2}{*}{Symbol} & \multirow{2}{*}{Description} & \multirow{2}{*}{Unit} & NGC5166 & NGC5166 & NGC5166 & NGC5166 \\
	    &&&& (fixed $r_\text{bulge}$) & (fixed $r_\text{bulge}$) & (with mock data) \\
	    \hline
		 $\theta$ & Inclination angle & - & 86.5 & 82.9 & 81.0 & 77.6 \\
		 $\Sigma_\text{dust}$ & Dust column density & $\text{M}_\odot/\text{kpc}^2$ & $8.79 \times 10^{4}$ & $1.53 \times 10^{5}$ & $1.94 \times 10^{5}$ & $2.01 \times 10^{5}$ \\
		 $r_\text{dust}/r_\text{disk}$ & Dust disk radius scaled by stellar disk radius & - & 4.34 & 2.68 & 3.39 & 4.64 \\
		 $r_\text{disk}$ & Stellar disk radius & kpc & 4.18 & 4.91 & 3.29 & 2.76 \\
		 $r_\text{bulge}$ & Stellar bulge radius & kpc & 0.32 & 0.64 & 3.17 & 1.85 \\
		 $\eta^\text{tot}$ & Intrinsic bolometric luminosity & $\text{L}_\odot$ & $9.91 \times 10^{10}$ & $8.04 \times 10^{10}$ & $7.94 \times 10^{10}$ & $6.97 \times 10^{10}$ \\
		 $\alpha_\text{B/T}$ & Intrinsic bulge to total luminosity ratio & - & 0.28 & 0.27 & 0.05 & 0.10 \\
		 $c^\text{disk}_1$ & Fractional luminosity for 1 Myr - 24 Myr & - & 0.00 & 0.00 & 0.28 & 0.08 \\
		 $c^\text{disk}_2$ & Fractional luminosity for 24 Myr - 581 Myr & - & 0.00 & 0.00 & 0.00 & 0.00 \\
		 $c^\text{disk}_3$ & Fractional luminosity for 581 Myr - 14 Gyr & - & 1.00 & 1.00 & 0.72 & 0.92 \\
		 $Z_\text{disk}$ & Disk metallicity & $\text{Z}_\odot$ & 0.43 & 1.33 & 1.33 & 1.33 \\
		 $c^\text{bulge}_1$ & Fractional luminosity for 1 Myr - 24 Myr & - & 1.00 & 1.00 & 0.00 & 1.00 \\
		 $c^\text{bulge}_2$ & Fractional luminosity for 24 Myr - 581 Myr & - & 0.00 & 0.00 & 0.00 & 0.00 \\
		 $c^\text{bulge}_3$ & Fractional luminosity for 581 Myr - 14 Gyr & - & 0.00 & 0.00 & 1.00 & 0.00 \\
		 $Z_\text{bulge}$ & Bulge metallicity & $\text{Z}_\odot$ & 1.33 & 0.002 & 1.33 & 0.002 \\
		\hline
	\end{tabular}}
\end{table*}

\begin{table*}
    \movetableright -10mm
	\centering
	\caption{Summary of some relevant galaxy properties derived from the best-fitting parameters of IC4225 and NGC5166.}
	\label{tab:compare}
	\begin{tabular}{cccccccc} 
		\hline
	    \multirow{2}{*}{Symbol} & \multirow{2}{*}{Description} & \multirow{2}{*}{Unit} & IC4225 & IC4225 & IC4225 & IC4225 & IC4225 \\
	    &&&& (fixed $r_\text{bulge}$) & (fixed $r_\text{bulge}$) & (with mock data) & \citepalias{2015MNRAS.451.1728D}$^\text{c}$ \\
		\hline
		 $\theta$ & Inclination angle & - & 88.7 & 86.8 & 86.6 & 86.5 & 89.3 \\
		 $r_\text{disk}$ & Stellar disk radius & kpc & 8.28 & 2.06 & 1.96 & 1.89 & 3.38 \\
		 $r_\text{bulge}$ & Stellar bulge radius & kpc & 0.32 & 0.64 & 1.25 & 1.41 & 1.25 \\
		 $r_\text{dust}$ & dust disk radius & kpc & 6.86 & 8.81 & 8.31 & 8.13 & 10.01 \\
		 $m_\text{dust}$ & Dust mass & $\text{M}_\odot$ & $3.05 \times 10^{7}$ & $3.28 \times 10^{7}$ & $3.02 \times 10^{7}$ & $2.98 \times 10^{7}$ & $2.10 \times 10^{7}$ \\
		 $m_*$ & Stellar mass$^\text{a}$ & $\text{M}_\odot$ & $6.64 \times 10^{10}$ & $6.90 \times 10^{10}$ & $6.84 \times 10^{10}$ & $6.86 \times 10^{10}$ & $3.89 \times 10^{10}$ \\
		 $m_\text{disk}$ & Disk Mass$^\text{a}$ & $\text{M}_\odot$ & $1.41 \times 10^{7}$ & $6.86 \times 10^{10}$ & $6.84 \times 10^{10}$ & $6.86 \times 10^{10}$ &  -  \\
		 $m_\text{bulge}$ & Bulge Mass$^\text{a}$ & $\text{M}_\odot$ & $6.64 \times 10^{10}$ & $3.25 \times 10^{8}$ & $8.79 \times 10^{6}$ & $8.49 \times 10^{6}$ &  -  \\
		 $\text{SFR}$ & Total star formation rate$^\text{b}$ & $\text{M}_\odot/\text{yr}$ & 0.12 & 0.46 & 0.07 & 0.07 & 0.26 \\
		 $\text{SFR}_\text{disk}$ & Disk star formation rate$^\text{b}$ & $\text{M}_\odot/\text{yr}$ & 0.12 & 0.00 & 0.00 & 0.00 &  -  \\
		 $\text{SFR}_\text{bulge}$ & Bulge star formation rate$^\text{b}$ & $\text{M}_\odot/\text{yr}$ & 0.00 & 0.46 & 0.07 & 0.07 &  -  \\
		 $\xi^\text{tot}_\text{abs}$ & Dust absorption fraction & - & 0.12 & 0.11 & 0.11 & 0.11 &  -  \\
		\hline
	    \multirow{2}{*}{Symbol} & \multirow{2}{*}{Description} & \multirow{2}{*}{Unit} & NGC5166 & NGC5166 & NGC5166 & NGC5166 & NGC5166 \\
	    &&&& (fixed $r_\text{bulge}$) & (fixed $r_\text{bulge}$) & (with mock data) & \citepalias{2015MNRAS.451.1728D}$^\text{c}$ \\
		\hline
		 $\theta$ & Inclination angle & - & 86.5 & 82.9 & 81.0 & 77.6 & 87.6 \\
		 $r_\text{disk}$ & Stellar disk radius & kpc & 4.18 & 4.91 & 3.29 & 2.76 & 3.92 \\
		 $r_\text{bulge}$ & Stellar bulge radius & kpc & 0.32 & 0.64 & 3.17 & 1.85 & 3.17 \\
		 $r_\text{dust}$ & dust disk radius & kpc & 18.11 & 13.15 & 11.15 & 12.79 & 5.81 \\
		 $m_\text{dust}$ & Dust mass & $\text{M}_\odot$ & $1.81 \times 10^{8}$ & $1.66 \times 10^{8}$ & $1.51 \times 10^{8}$ & $2.07 \times 10^{8}$ & $4.80 \times 10^{7}$ \\
		 $m_*$ & Stellar mass$^\text{a}$ & $\text{M}_\odot$ & $1.11 \times 10^{11}$ & $1.10 \times 10^{11}$ & $1.10 \times 10^{11}$ & $1.09 \times 10^{11}$ & $6.61 \times 10^{10}$ \\
		 $m_\text{disk}$ & Disk Mass$^\text{a}$ & $\text{M}_\odot$ & $1.11 \times 10^{11}$ & $1.10 \times 10^{11}$ & $1.02 \times 10^{11}$ & $1.09 \times 10^{11}$ &  -  \\
		 $m_\text{bulge}$ & Bulge Mass$^\text{a}$ & $\text{M}_\odot$ & $1.08 \times 10^{8}$ & $4.89 \times 10^{7}$ & $7.44 \times 10^{9}$ & $1.52 \times 10^{7}$ &  -  \\
		 $\text{SFR}$ & Total star formation rate$^\text{b}$ & $\text{M}_\odot/\text{yr}$ & 0.92 & 0.42 & 0.69 & 0.29 & 0.93 \\
		 $\text{SFR}_\text{disk}$ & Disk star formation rate$^\text{b}$ & $\text{M}_\odot/\text{yr}$ & 0.00 & 0.00 & 0.69 & 0.16 &  -  \\
		 $\text{SFR}_\text{bulge}$ & Bulge star formation rate$^\text{b}$ & $\text{M}_\odot/\text{yr}$ & 0.92 & 0.42 & 0.00 & 0.13 &  -  \\
		 $\xi^\text{tot}_\text{abs}$ & Dust absorption fraction & - & 0.21 & 0.25 & 0.24 & 0.24 &  -  \\
		 \hline
	\end{tabular}
	\begin{tablenotes}
	    \item $^\text{a}$ The disk mass and bulge mass are obtained by integrating the star formation history without taking into account the mass loss due to the supernova. 
	    \item $^\text{b}$ The timescale of star formation rate is $\sim 100 \, \text{Myr}$.
	    \item $^\text{c}$ \cite{2015MNRAS.451.1728D} estimated the stellar mass and star formation rate using \textsc{magphys} \citep{2008MNRAS.388.1595D} and the other properties using \textsc{fitskirt} \citep{2013A&A...550A..74D}. 
	\end{tablenotes}
\end{table*}

\begin{figure*}[h]
	\includegraphics[width=\textwidth]{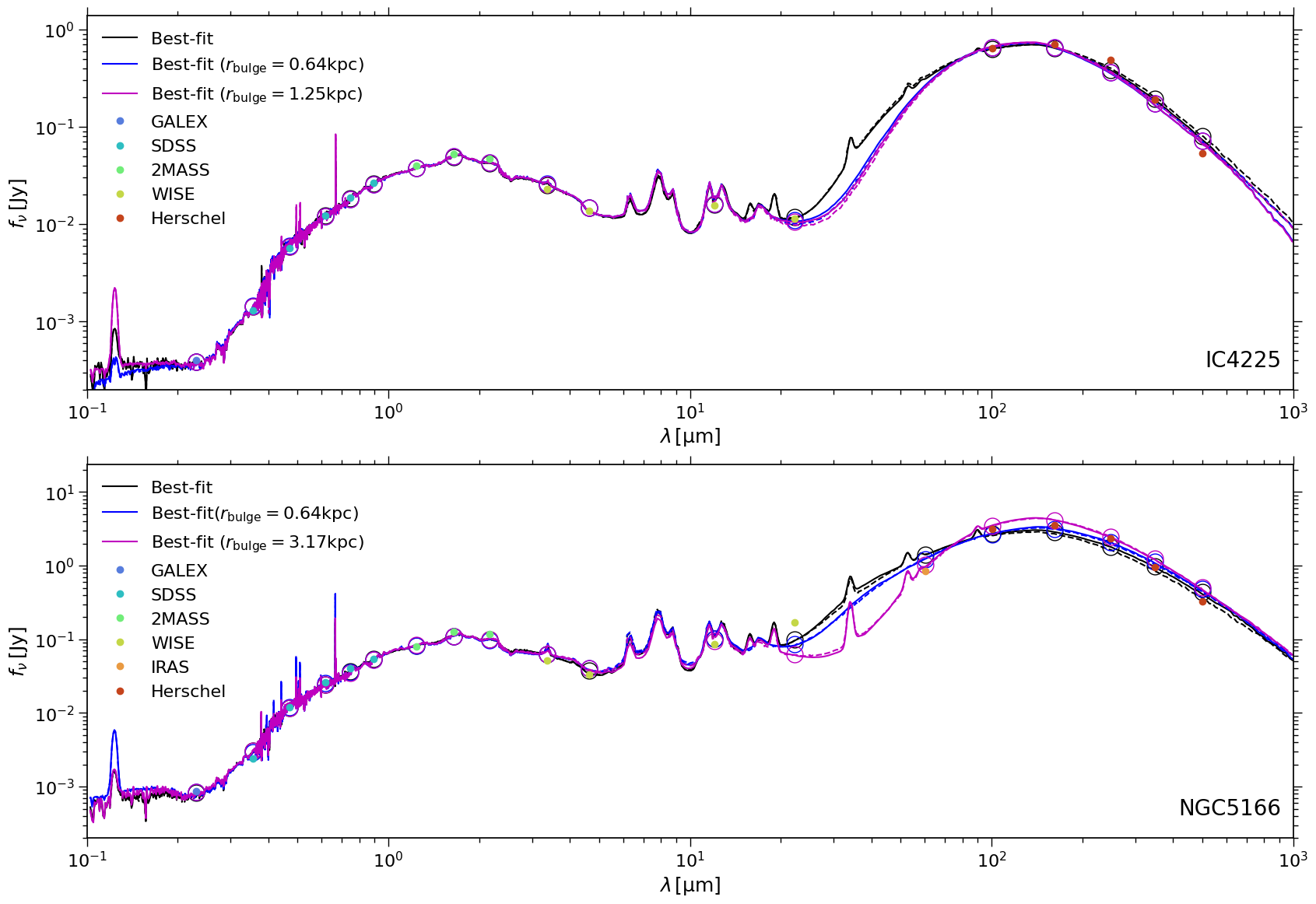}
    \caption{Best-fitting SEDs of IC4225 and NGC5166. Coloured dots are observational data described in \citet{2015MNRAS.451.1728D}. Solid lines with empty circles are the best-fitting results of our deep learning SED model. For the black lines, no parameter is fixed, while for the blue and magenta lines, bulge radius is fixed. In particular, when fitting the magenta lines, we fix the bulge radius at the same value as \citet{2015MNRAS.451.1728D}. The dashed lines are obtained by running the \textsc{skirt} simulation using the corresponding best-fitting parameters listed in Table \ref{tab:best-fit}. The model free parameters and the fitting method are described in Section \ref{sec:app}. As discussed in Section \ref{sec:app}, observational uncertainties are not considered in the fittings, which may be smaller than the intrinsic errors of our deep learning model.}
    \label{fig:best_fit}
\end{figure*}

\begin{figure*}[h]
	\includegraphics[width=\textwidth]{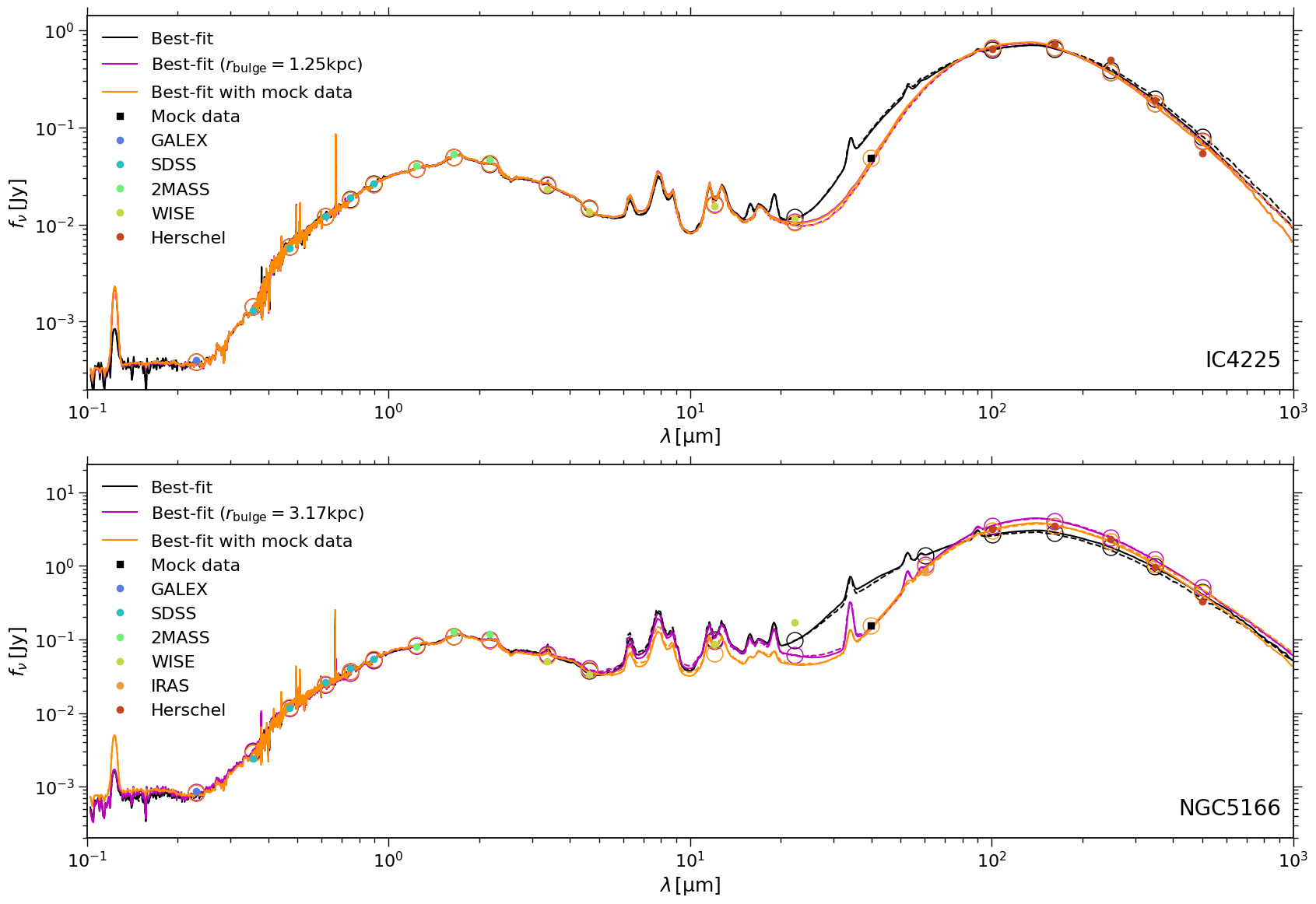}
    \caption{Best-fitting SEDs of IC4225 and NGC5166 with mock data at $40 \micron$, which are shown as orange lines. The mock data is computed using the best-fitting SEDs with fixed bulge radius, which are shown as magenta lines. For IC4225, the orange line overlaps the magenta line. The back lines correspond to the best-fitting results without fixed parameters. The dashed lines are obtained by running the \textsc{skirt} simulation using the corresponding best-fitting parameters listed in Table \ref{tab:best-fit}. The model free parameters and the fitting method are described in Section \ref{sec:app}. As discussed in Section \ref{sec:app}, observational uncertainties are not considered in the fittings, which may be smaller than the intrinsic errors of our deep learning model.}
    \label{fig:best_fit_extra}
\end{figure*}

\section{Applications} \label{sec:app}
As an application, we fit our deep learning model to the observed SED of IC4225 and NGC5166. We use the data summarised by \citet{2015MNRAS.451.1728D} (hereafter \citetalias{2015MNRAS.451.1728D}), which are obtained from the \textit{GALEX} \citep{2005ApJ...619L...1M}, the SDSS \citep{2009ApJS..182..543A}, the 2MASS \citep{2006AJ....131.1163S}, the \textit{WISE} \citep{2010AJ....140.1868W}, the \textit{IRAS} \citep{1992ifss.book.....M}, and the \textit{Herschel}-ATLAS  \citep{2010PASP..122..499E}. The used data are plotted in Figure \ref{fig:best_fit}.
\par
For the present work, we do not adopt the uncertainties from the observations. The observed errors are assumed to be Gaussian in flux. However, as discussed in Section \ref{sec:valid}, our deep learning model also introduces some uncertainties, which are Gaussian-like in magnitude. Some of these errors may be even larger than the observations. In addition, the \textsc{skirt} simulation itself also introduces $\sim 0.05$ mag numerical errors as suggested by \cite{2018ApJS..234...20C}. Hence, for the fittings of IC4225 and NGC5166, we employ a loss function that assumes Gaussian errors in magnitude.
\begin{equation} 
    \Psi = \sum_\text{band} \left( \frac{m^\text{obs}_\text{AB} - m^\text{true}_\text{AB}}{\sigma} \right)^2.
    \label{eqn:loss_fit}
\end{equation}
Based on the results illustrated in Figure \ref{fig:valid_errors}, we assume $\sigma = 0.1$ for the \textit{WISE}-12, \textit{WISE}-22, and \textit{IRAS}-100 bands, and $\sigma = 0.2$ for all \textit{Herschel} bands. For the other bands, we adopt $\sigma = 0.05$ to take into account the errors from the \textsc{skirt} simulation. In addition, according to the left panel of Figure \ref{fig:valid_errors}, we also add a $\sim -0.006$ mag offset to those bands. However, we find that this correction has a small influence on the results. Since we ignore the observational errors, to avoid double weighting, we do not use the fluxes in \textit{IRAS}-25, \textit{IRAS}-100 and \textit{Planck} bands for NGC5166, which overlap the existing data points. In addition, we assume the distances of IC4225 and NGC5166 are 73.9 Mpc and 62.6 Mpc respectively according to \citetalias{2015MNRAS.451.1728D}.
\par
The free parameters of the fitting model are the input parameters of our deep learning model. As listed in Table \ref{tab:params}, we have six parameters describing the geometry and two parameters specifying the intrinsic bolometric luminosity of the stellar disk and bulge. As introduced in Section \ref{sec:nn_de}, we can have thirty-six parameters for the star formation history. Such degrees of freedom, however, may be unnecessary for the SED fitting. Instead, we use three parameters to describe the fractional luminosity of three stellar age bins and one parameter to specify the overall metallicity. The reduced stellar age bins are obtained by merging every two bins. The corresponding bin edges can be found in Table \ref{tab:best-fit}. The metallicity is applied using linear interpolation. This is a typical discrete star formation history model used by several recent SED-fitting studies \citep{2017ApJ...837..170L,2019A&A...622A.103B,2020MNRAS.495..905R}. We adopt this model for both the stellar disk and bulge. We note that the fractional luminosity parameters $c_k$ should satisfy
\begin{equation}
    \sum_k c_k = 1
\end{equation}
To tackle this constrain, we assume that each input parameter satisfies $0 < a_k < 1$, and apply the following transformation
\begin{align}
    & c_k = \frac{\log(1 - a_k)}{\sum_k \log(1 - a_k)}
\end{align}
If each $a_k$ is uniformly distributed, the transformed parameters will follow a flat Dirichlet distribution. Overall, our SED-fitting model contains 15 free parameters.
\par
We utilise the particle swarm optimisation (PSO) \citep{shi1998modified} to minimise the loss function defined by Equation (\ref{eqn:loss_fit}). We run the optimiser with 32 particles for 2000 iterations. We find that the parameter space is complex and contains multiple local minima. To ensure the results are stable, we perform the optimisation 10 times for each galaxy, and choose the best-fit from these runs.
\par
The best-fitting SEDs are demonstrated in Figure \ref{fig:best_fit}, and the best-fitting parameters are given in Table \ref{tab:best-fit}. We find that our SED model can fit the observations well. To test our deep learning model, we run the \textsc{skirt} simulation using the best-fitting parameters, and plot the results in Figure \ref{fig:best_fit} as dashed lines. The SEDs predicted by our deep learning model are well consistent with the simulation results.
\par
 Table \ref{tab:compare} compares some relevant quantities with \citetalias{2015MNRAS.451.1728D}, who carried out an image fitting analysis on these galaxies using \textsc{fitskirt} \citep{2013A&A...550A..74D}. Our model also predicts that both IC4225 and NGC5166 are edge-on galaxies, consistent with the galaxy images. Our predicted galaxy properties are broadly consistent with \citetalias{2015MNRAS.451.1728D} but with some anomalies. For IC4225, the stellar disk radius is overestimated by a factor of 2. For NGC5166, our results overestimate the dust disk radius and the dust mass by a factor of 3 and 4 respectively. An additional issue is that for NGC5166, our model cannot fit the fluxes in the \textit{WISE}-22 and \textit{IRAS}-70 bands. This behaviour was also found by \citetalias{2015MNRAS.451.1728D}. Their results show even poorer consistency with the observations at MIR to FIR wavelengths. The issue is known as the dust energy balance problem, which was reported by several image studies \citep[e.g.][]{2012MNRAS.427.2797D,2016A&A...592A..71M,2018A&A...616A.120M}. Moreover, for both galaxies, the predicted bulge radius is significantly underestimated and reaches the minimum value in our model\footnote{We note that our geometry model of the stellar bulge is slightly different from \citetalias{2015MNRAS.451.1728D}, whose model includes two additional parameters: the S\'ersic index and the intrinsic flattening.}.
\par
To understand the discrepancies, we examine the impact of bulge radius on predicted SEDs. To examine the dependency in the SED on bulge radius, we perform two additional fittings for each galaxy, doubling the bulge radius and fixing the bulge radius at the same value with \citetalias{2015MNRAS.451.1728D}. The additional results can also be found in Figure \ref{fig:best_fit}, Tables \ref{tab:best-fit} and \ref{tab:compare}. Figure \ref{fig:best_fit} shows that increasing the bulge radius lowers the fluxes at $20 \, \micron - 80 \, \micron$. This trend is very obvious for NGC5166, in which case the predicted SED has a constraint at $70 \, \micron$. In Tables \ref{tab:best-fit} and \ref{tab:compare}, while no apparent trend can be found in the changes of the best-fitting parameters, the results become more consistent with \citetalias{2015MNRAS.451.1728D}, particularly for IC4225. The relative sizes of the dust disk, stellar disk and bulge can affect the dust optical depth significantly, resulting in different behaviours of dust attenuation and emission. Our results imply that these parameters are highly degenerated.
\par
We propose that including information at $20 \, \micron - 80 \, \micron$ could alleviate the degeneracy in the geometry parameters. To verify this, we carry out a fitting for each galaxy with mock data at $40 \, \micron$. The mock data are estimated from the best-fitting results of using the same bulge radius as \citetalias{2015MNRAS.451.1728D}. We use a tophat filter centred at $40 \, \micron$ with a width of $10 \, \micron$ to compute the flux. In addition, for NGC5166, we treat the flux in \textit{WISE}-22 as an outlier and exclude it in the fitting. We show the additional results in Figure \ref{fig:best_fit}, Tables \ref{tab:best-fit} and \ref{tab:compare}. When including the mock data at $40 \, \micron$, we obtain better constraints for the bulge radius and achieve better consistency with \citetalias{2015MNRAS.451.1728D} for the geometry parameters. In particular, for IC4225, the results with mock data and fixed bulge radius are quite consistent. As illustrated in Figure \ref{fig:best_fit_extra}, the obtained SEDs in both cases overlap completely. The additional fittings present above imply that fluxes at $20 \, \micron - 80 \, \micron$ are related to bulge radius and could reduce the degeneracy in the parameter space of our SED model. However, this point should be tested using a large galaxy sample. These fluxes may be observed by \textit{Herschel} in galaxies at higher redshifts.
\par
Some implications can also be found by comparing the estimated stellar mass and dust mass among the different fittings of each galaxy. We have carried out four different fittings for both IC4225 and NGC5166, leaving all parameters free, fixing the bulge radius and including mock data. As demonstrated in Table \ref{tab:params}, while the variations in stellar mass, dust mass and dust absorption fraction are small, geometry parameters such as dust disk radius show large differences. Our predicted stellar mass is slightly overestimated compared with \citetalias{2015MNRAS.451.1728D}, who measured the stellar mass using \textsc{magphys} \citep{2008MNRAS.388.1595D}. This difference is consistent with the study by \cite{2021ApJ...923...26D}, who found that inclination independent SED models could underestimate stellar mass. In addition, the different choices of the simple stellar population library could also impact the measurements of the stellar mass. In terms of the dust mass, while our predicted value for IC4225 is broadly consistent with \citetalias{2015MNRAS.451.1728D}, there is a disagreement for NGC5166. A potential reason could be that both \textsc{fitskirt} and our SED model cannot reproduce the flux in \textit{WISE}-22, which introduces systematic errors in the fittings.

\section{Summary} \label{sec:summary}
In this work, we present a deep learning model which predicts the FUV to FIR SED by emulating the \textsc{skirt} radiative transfer simulation. The SED model is comprised of three specifically designed neural networks. Each neural network can be trained using $\sim 4000 - 5000$ samples. We estimate that the uncertainties of the deep learning model are $\sim 0.005$ for the dust attenuation neural network and $\sim 0.1 - 0.2$ for the dust emission neural network.
\par
As an application, we use our deep learning SED model to fit the observed SEDs of IC4225 and NGC5166, and compare the results with the image-fitting analysis by \citet{2015MNRAS.451.1728D}. Our findings can be summarised as follows:
\begin{enumerate}
    \item Our model can reproduce the observed SEDs, and provide reasonable estimations of the inclination angle and stellar mass.
    \item The predicted fluxes at $20 \, \micron - 80 \, \micron$ by our SED model are correlated with bulge radius.
    \item The parameter space of our SED model is highly degenerated. We find that including mock data at $\sim 40 \, \micron$ could reduce the degeneracy, in which case our predicted disk radius and bulge radius are broadly consistent with the image-fitting analysis.
\end{enumerate}
\par
Our model can also be applied to generate SEDs for simulated galaxies. Our model includes no phenomenological parameters such as attenuation curve slope, which avoids replying on empirical relations. As input to our SED model, disk radius, bulge radius and star formation history are direct predictions of semi-analytic models, while dust mass could be obtained by an additional dust evolution model \citep[e.g.][]{2017MNRAS.471.3152P,2019MNRAS.489.4072V,2021MNRAS.503.1005T} or using a dust-to-total metal ratio motivated by observations \citep[e.g.][]{2016MNRAS.462.3854L,2019MNRAS.489.4196L}. Moreover, dust radius is an unknown parameter, which requires further modelling. In future work, we will apply our SED model to a semi-analytic model.
\par
This work demonstrates the feasibility of using the supervised deep learning technique to accelerate the speed of radiative transfer simulations. We expect that our model can be widely used in SED-fitting and SED-modelling of galaxies from semi-analytic models. Generating theoretical galaxy SEDs is a very complex task, and there are still many aspects that are not considered in this work. We suggest that our SED model can have the following extensions
\begin{enumerate}
    \item It is instructive to have the SED model support multiple stellar evolutionary models. While our dust attenuation model is independent of simple stellar population (SSP) libraries, our dust emission model is restricted to the Flexible Stellar Population Synthesis library with the given star formation history bins, initial mass function and nebular emission model. If we want to change the settings, we need to run additional radiative simulations and retrain the model.
    \item Our dust attenuation model on emission lines can be improved. The attenuation from the birth cloud of young stars should be taken into account explicitly. This is critical for spectroscopic and emission line studies.
    \item The uncertainties of our dust emission model are not negligible for applications of SED-fitting. This issue might be resolved by improving the design of the neural network. Alternatively, one can also build a model to describe the errors.
\end{enumerate}

\begin{acknowledgments}
This work is partly supported by the NSFC (No.11825303, 11861131006), the science research grants from the China Manned Space project with NO.CMS-CSST-2021-A03, CMS-CSST-2021-B01 and the cosmology simulation database (CSD) in the National Basic Science Data Center (NBSDC-DB-10). We thank the reviewer for providing a constructive report, which improves the quality of the paper. We acknowledgement a helpful discussion with Peter Camps and Maarten Baes.
\end{acknowledgments}

%



\software{
    \textsc{astropy}\footnote{https://www.astropy.org/} \citep{astropy:2013, astropy:2018}, \textsc{fsps} \citep{2009ApJ...699..486C,2010ApJ...712..833C}, \textsc{jupyter-notebook}\footnote{https://github.com/executablebooks/jupyter-book}, \textsc{matplotlib} \citep{2007CSE.....9...90H}, \textsc{numpy} \citep{harris2020array}, \textsc{pytorch} \footnote{https://pytorch.org/} \citep{2019arXiv191201703P}, \textsc{python-fsps}\footnote{https://dfm.io/python-fsps/current/}, \textsc{scipy} \citep{scipy}, \textsc{sedpy} \footnote{https://github.com/bd-j/sedpy}, \textsc{skirt} \citep{2020A&C....3100381C}
}





\bibliography{references}{}
\bibliographystyle{aasjournal}



\end{document}